\documentclass[12pt]{article}

\input epsf
\def\beq{\begin{eqnarray}}
\def\eeq{\end{eqnarray}}

\def\lsim{\mathrel{\rlap{\lower3pt\hbox{\hskip0pt$\sim$}}
     \raise1pt\hbox{$<$}}}         
\def\gsim{\mathrel{\rlap{\lower4pt\hbox{\hskip1pt$\sim$}}
     \raise1pt\hbox{$>$}}}         

\begin{document}


\begin{flushright}
{ NYU-TH-04/11/6}
\end{flushright}
\vskip 0.9cm

\centerline{\Large \bf 
On Theta Dependence of
Glueballs from AdS/CFT}

\vskip 0.7cm
\centerline{\large Gregory Gabadadze and Alberto Iglesias}
\vskip 0.3cm
\centerline{\em Center for Cosmology and Particle Physics}
\centerline{\em Department of Physics, New York University, New York, 
NY, 10003, USA}

\vskip 1.9cm 

\begin{abstract}

We study the theta dependence of the glueball spectrum in a 
strongly coupled cousin of large $N$ gluodynamics defined via the 
AdS/CFT correspondence. By explicitly diagonalizing the 10d 
gravity equations in the presence of the RR 3-form and 1-form fluxes
we found a  mixing pattern for the lowest-spin lightest glueballs. 
The mixing between the scalar and  pseudoscalar states
is not suppressed, suggesting that the CP-odd effects 
persist in the large $N$ theory.  As a consequence, 
the lightest mass eigenstate  ceases to be a parity eigenstate. We 
found the former as a linear combination  of a scalar and 
pseudoscalar glueballs. On the other hand, the mass eigenvalues in a 
theory with and without the theta term remain equal in the large $N$ limit.

\end{abstract}

\newpage

\section{Introduction and summary}

QCD in the limit of a large number of colors, $N \to \infty$, is expected 
to contain a great deal of information on confinement \cite {tHooft}.  
In this limit gluons dominate the partition function and 
confinement should be easier to study in the theory without quarks.
The subject of the present work is large $N$ gluodynamics
(i.e., QCD without quarks) with a nonzero theta term. 
Observationally, the theta parameter, if not zero, should be truly tiny, 
$\theta<10^{-9}$, and, studies of  the theta dependence in the large $N$ 
theory  may seem less motivated. However, this is not 
so because of the following arguments. Large $N$ gluodynamics with the theta 
term reveals a rich ground state structure with multiple vacua 
separated by domain walls  \cite{Witten1,Witten:1998uk,Shifman,GG}.  
This by itself is  an extremely interesting  manifestation of nonperturbative 
physics in a non-Abelian gauge theory. Moreover, 
certain remnants of the large $N$ vacuum structure 
are expected to be present in real QCD too. Examples of this include  
the heavy nonperturbative states that make the domain walls \cite {GSPR},
and new properties of axion and hadronic domain walls in a theory where  
the strong CP problem is solved by the axion mechanism (for recent summary  
see Ref. \cite {GS}). 

The theta term enters the QCD action suppressed by one power of $N$
compared to the gluon kinetic term.  This may  
seem to indicate that no theta dependence should survive
in the large $N$ limit. For instance, a dilute instanton gas 
approximation would give rise to the theta dependence that is 
suppressed as ${\rm exp}(-N)$. However, there is a  
body of evidence (lattice and otherwise) suggesting that this is 
not so. The best example is the theta dependence of the vacuum energy 
\cite {Veneziano,Witten1,Ohta,Witten:1998uk} which is of order ${\cal O}(N^0)$.
This dependence can only be attributed to certain  
infrared effects that are not captured by a dilute 
instanton gas approximation. The related issue that has not been explored 
so far is the theta dependence of the spectrum of glueballs in this theory.
Masses of glueballs are generated by noperturbative 
effects as well \cite {Alike}. Similar effects could in general 
introduce the theta dependence in the spectrum even  
in the large $N$ limit.  In particular, we did not 
find convincing arguments to believe that the theta dependence would be  
completely washed out from the spectrum in the large $N$ limit. 
How would one study these issues in more detail? 

The theta term introduces a complex phase in the 
euclidean formulation of the partition function of 
a theory the action of which is otherwise 
real and positive semidefinite. Because of this
lattice studies of these issues seem difficult (as an exception, see, 
{\it e.g.}, \cite{vicari}).
One way to address the question of the theta dependence of glueballs is to use 
the AdS/CFT correspondence \cite {Maldacena,Klebanov,Witten:1998qj} 
adapted to nonsupersymmetric and non-conformal theories  
(for a review see, \cite {adscftrev}). A well-known 
drawback of this approach is that it only allows to calculate 
gauge theory observables for a strongly coupled cousin of  large $N$ 
gluodynamics which in addition contains states that are not present in 
QCD. However, in the large $N$ limit the effects of the additional 
states on QCD resonance physics should be minimized, and in the absence 
of other methods this seems to be a reasonable starting place.

\newpage

In the main part of the work we study the theta dependence of the glueballs  
by looking at the $IIA$ construction of $N$ D4-branes with 
one compact worldvolume dimension
\cite{Witten:1998zw} and  a nonzero bulk $RR$ 1-form flux.  
The latter gives rise to the theta dependence in the dual 
gauge theory \cite {Witten:1998uk}. In this setup we manage to 
diagonalize exactly the bulk gravity equations for the fluctuations 
of a graviton, dilaton  and RR 1-form. The diagonalization of the 
gravity equations suggest the following 
remarkable pattern for the theta dependence of the gauge theory glueballs.
The lowest  scalar $(0^{++})$ and 
pseudoscalar $(0^{-+})$ glueballs  mix. 
There are  kinetic as well as
mass mixing terms both of which are determined by $\theta$.
The system of glueballs can be diagonalized by 
shifting the field of  the lightest spin-zero state
by a field that is proportional to the heavier spin-zero 
glueball multiplied by $\theta$.  Interestingly enough, 
the above diagonalization can only be achieved by shifting the 
field  of the lightest spin-zero state, which in gluodynamics is necessarily 
the $0^{++}$ glueball \cite {West}.  The heavier state remains 
intact. This property, as we will discuss in Section 4, is vital
for the applicability of the method of the calculation. 
After the diagonalization, the lowest mass eigenstate ceases to be  
a parity eigenstate. The mixing between the parity eigenstates 
is a leading effect in large $N$ (i.e., it is not suppressed by powers of 
$N$). However, the masses of the physical (diagonal) states in a theory 
with and without the theta term are the same. Hence, the only 
effect of the  theta term in the leading large $N$ limit is that 
the lowest mass spin-zero state becomes a mixed state of $0^{++}$ 
and $0^{-+}$ glueballs. The periodicity of the wavefunction with respect to 
$\theta \to \theta + 2\pi ({\rm integer})$  
is achieved by choosing appropriate branches of the theory in a way that 
also makes the vacuum energy periodic in $\theta$ \cite 
{Witten1,Witten:1998uk}.


\section{$D4$ Soliton with $RR$ $1-$form flux}

{}In order to study the theta dependence of glueballs from the Gauge/Gravity 
correspondence we consider the dual supergravity description of pure $U(N)$ 
gauge field theory introduced in \cite{Witten:1998zw}. In this setup one starts
 with weakly coupled type $IIA$ superstring theory in the presence of $N$ 
$D4$-branes. The $D4$ {\it soliton} background solution relevant for this discussion 
is obtained 
by compactifying on a circle of radius $M_{KK}^{-1}$ one of the 4 spacial 
directions of the 4-brane solution and by taking the near 
horizon limit.  The effect of the compactification in the 
$D4$-branes worldvolume theory is that the fermions acquire 
masses of order $M_{KK}$  
at tree level due to the  anti-periodic boundary conditions and so do the 
scalars due to loops. The worldvolume theory at energies below 
$M_{KK}$ is then 
4-dimensional nonsupersymmetric nonconformal $U(N)$ gauge theory. Thus we 
proceed to review the $D4$ soliton solution and how the effect of a theta term 
is incorporated.
 
{}The low energy effective action of type $IIA$ superstrings in the Einstein 
frame and with vanishing Kalb-Ramond $B$ field reduces to
\begin{equation}
I={1\over (2\pi)^7 l_s^8}
\int {\rm d}^{10}x\sqrt{g}\left[g_s^{-2}\left({\cal R}-
{1\over 2}\partial_M\phi\partial^M\phi\right)
-{1\over 4}{\rm e}^{3\phi/2}F_{(2)}^2
-{1\over 48}{\rm e}^{\phi/2}F_{(4)}^2\right]~,
\end{equation}
where $l_s$ and $g_s$ are the string length and coupling constant and $F_{(2)}$
 and $F_{(4)}$ 
are the field strengths of the $RR$ 1 and 3-forms respectively. From these, 
the equations of motion for the metric, dilaton and 1-form follow:
\begin{eqnarray}
g_s^{-2}\left({\cal R}_{MN}-{1\over 2}\partial_M\phi\partial_N\phi\right)&=&
{1\over 2}{\rm e}^{3\phi/2}\left(F_M^{~P}F_{NP}-{1\over 16}g_{MN}F_2^2\right)-
\nonumber \\ 
&&~~-{1\over 48}{\rm e}^{\phi/2}\left(4F_{MRST}F_N^{~RST}-{3\over 8}g_{MN}
F_{(4)}^2
\right)~,\label{geom}\\
g_s^{-2}{1\over\sqrt{g}}\partial_M\left(\sqrt{g}g^{MN}\partial_N\phi\right)&=&
{3\over 8}{\rm e}^{3\phi/2}F_{(2)}^2+{1\over 48}{\rm e}^{\phi/2}F^2_{(4)}~,
\label{phieom}\\
\partial_M\left(\sqrt{g}{\rm e}^{\phi/2}F^{MN}\right)&=&0~. \label{feom}
\end{eqnarray}
The so called $D4$ {\it soliton} is given by the following Einstein frame 
metric, dilaton and constant 4-form: 
\begin{eqnarray}
{\rm d}s^2&=&\left({R\over U}\right)^{3/8}\left[\left({U\over R}\right)^{3/2}
\left(f(U){\rm d}\tau^2+{\rm d}x^\mu{\rm d}x_\mu\right)+
\left({R\over U}\right)^{3/2}{{\rm d}U^2\over f(U)}+
R^{3/2}U^{1/2}{\rm d}\Omega^2_4\right]~,\\
{\rm e}^{\phi}&=&\left({U\over R}\right)^{3/4}~, \hskip2cm 
F_{(4)}~\sim~ g_s^{-1}\epsilon_4~, \nonumber\\
R^3&=&\pi g_sNl_s^3~,\hskip2cm f(U)~=~1-\left(U_{KK}/U\right)^3~,
\end{eqnarray}
where $x^\mu,~\mu=1\dots 4$ are the euclidean coordinates of the noncompact 
directions and $\tau$ denotes the coordinate  along the compactified circle 
on the $D4$ worldvolume.
${\rm d}\Omega_4^2$ and $\epsilon_4$ are 
the line element and volume form on a unit $4-$sphere and $U$ is the coordinate
on  the radial direction transverse to the $D4$ branes. In order to avoid 
a conical singularity at $U=U_{KK}$, 
$\tau$ is identified with period $\Delta\equiv 2\pi M_{KK}^{-1}
=4\pi R^{3/2}/3 U^{1/2}_{KK}$.

The supergravity description is valid in the regime of small curvatures (in 
string length units) and string coupling, namely 
\begin{equation}
l_s^2{\cal R}\ll 1~,~~~~~~~
g_s{\rm e}^\phi\ll 1~. 
\end{equation}
The latter implies a maximum value $U_{max}$ of the radial coordinate that 
should therefore be much larger than the parameter $U_{KK}$ 
\cite{Kruczenski:2003uq}.
This regime corresponds to the 't Hooft limit of the four dimensional theory
$g_{YM}\rightarrow 0$, $N\rightarrow \infty$, $g_{YM}^2N={\rm fixed}\gg 1$, 
where the identification used $g^2_{YM}\sim g_s l_s M_{KK}$ 
can be read off from
(\ref{bi4}) below.

{}The theta dependence of this background is obtained by turning on 
the $RR$ $1-$form $C_{(1)}$ in the $\tau$ direction 
\cite{Witten:1998uk}:
\begin{equation}\label{C}
C_{\tau}=-{U_{KK}^3\over \Delta}{\theta_c\over U^3}~,
\end{equation}
where $\theta_c=\theta+2 n \pi$ with integer $n$ and the normalization has been
chosen such that $\int{\rm d}\tau{\rm d}U F_{U\tau}=\theta_c$. 
Indeed, it is not difficult 
to see that (\ref{C}) solves (\ref{feom}) if the background geometry is not 
altered by the presence of this new field. This absence of backreaction on the 
geometry is due to the fact that in the large $N$ limit the contribution from 
$F_{(2)}$ to (\ref{geom}) and (\ref{phieom}) is subleading because the other 
terms contain extra powers of $g_s^{-1}\sim N$.


\section{Supergravity modes}

{}In order to find the glueball spectra, we proceed to study the equations of 
motion to linear order in the background of the previous section. 
In particular we are interested in 
the scalar modes that act as sources for the scalar and pseudoscalar operators
of the boundary theory. For the case of vanishing theta this modes have been 
identified in \cite{Hashimoto:1998if} and \cite{Constable:1999gb}. In this
 section we will 
find the particular combination that diagonalizes the supergravity equations of
 motion when theta is turned on.

{}For the metric and dilaton perturbations we consider the following 
linearization
\begin{equation}
g_{MN}=g^B_{MN}+h_{MN}~,~~~~~~~~\phi=\phi^B+\delta\phi~,
\end{equation}
where $g_{MN}^B$ and $\phi^B$ stand for the background values and the diagonal 
components of the fluctuations are parametrized as follows:
\begin{eqnarray}
h_{MM}&=&g^B_{MN} v^N H(U) {\rm e}^{i k_\mu x^\mu}~,\\
\delta\phi&=&\phi_0{\rm e}^{ik_\mu x^\mu} H(U)~.
\end{eqnarray}
Respecting the $SO(4)$ and $SO(5)$ symmetries of the $x^1\dots x^4$ and $S^4$ 
directions there are tree scalar fluctuations $T$, $L$ and $S$ (in
the notation of \cite{Brower:2000rp}) that correspond to the dilaton, 
$4-$sphere volume 
fluctuation and the exotic polarization 
\cite{Constable:1999gb} correspondingly. To simplify the expressions 
we choose $k^\mu=\delta^{\mu4}k_4$ via an $SO(4)$ rotation. These 
are given by:
\begin{eqnarray}
T:~~~v^N&=&{1\over 4}(1,-{5\over 3},-{5\over 3},-{5\over 3},1,1,1,1,1,1)~,
~~~\phi_0={3\over 2}~,\\
L:~~~v^N&=&(-1,-1,-1,-1,-1,-1,1,1,1,1)~,~~~\phi_0=-{2\over 3}~,\\
S:~~~v^N&=&{1\over 20}(-31,9,9,9,{-98+5U^3\over -2+5U^3},
{-98+5U^3\over -2+5U^3},1,1,1,1)~, ~~~\phi_0={3\over 10}.
\end{eqnarray}
The exotic polarization $S$ has also (in the gauge 
chosen) off diagonal components
\begin{equation}
h_{U x_4}=h_{x_4 U}=-i{k_4\over k^2}{72 U^{25/8}\over (5U^3-2)^2} H(U) 
{\rm e}^{i k_\mu x^\mu}~.
\end{equation}
For each of these polarizations, the function $H\equiv T,L,S$ respectively 
satisfies:
\begin{eqnarray}
&&U(U^3-1)T''+(4U^3-1)T'-k^2UT=0~,\\
&&U(U^3-1)L''+(4U^3-1)L'-(k^2U-18U)L=0~,\\
&&U(U^3-1)S''+(4U^3-1)S'-U\left(k^2-{108U\over (5U^3-2)^2}\right)S=0~,
\end{eqnarray}
where, for convenience, we have rescaled the coordinates to form dimensionless 
quantities, namely
\begin{equation}
\tilde \tau={U_{KK}^{1/2}\over R^{3/2}}\tau~,~~~~~~\tilde 
x^\mu={U_{KK}^{1/2}\over R^{3/2}}x^\mu~,~~~~~~\tilde U={U\over U_{KK}}~,
\end{equation}
and dropped the tildes.

{}Let us now turn our attention to the fluctuations of $C_\tau$. 
We linearize the equation of motion for the $RR$ 1-form (\ref{feom}) with 
fluctuations given by
\begin{equation}
C_\tau=C_\tau^B+\chi(U){\rm e}^{ik_\mu x^\mu}~,
\end{equation}
where $C_\tau^B$ is the background value. 

{}The key point is that the resulting equation can be decoupled 
from the metric and dilaton by choosing
\begin{equation}
\chi={3\over 2}V+{3\theta_c\over 2\pi}f(U)(T+S)~,
\end{equation}
with $V$ satisfying the following equation:
\begin{equation}
U(U^3-1)V''+4(U^3-1)V'-k^2U V=0~,
\label{V}
\end{equation}
the equation for fluctuations of $C_\tau$ in the absence of theta term. 
Thus, we were able to diagonalize the supergravity dual of $SU(N)$ gauge 
theory in the presence of a $CP$ violating term. This has important 
consequences that we discuss in the following section.


\section{Couplings to boundary theory}

{}We infer the coupling of the supergravity modes to gauge invariant operators 
of the four dimensional field theory by considering the Born-Infeld action 
describing the low energy worldvolume excitations of the $D4$ branes. The
operators of interest for us are $f_{\mu\nu}f^{\mu\nu}$ and 
$f_{\mu\nu}\tilde f^{\mu\nu}$ for Yang-Mills field-strength 
$f_{\mu\nu}$ and its dual $\tilde f^{\mu\nu}$. Neglecting 
worldvolume scalars it reads (in string frame):
\begin{equation}
I_{BI}={\cal T}_4{\rm Tr}\int {\rm d}^5x\left({\rm e}^{-\phi}\sqrt{{\rm det}
\left(g_{mn}+ 2\pi \alpha' f_{mn}\right)}+{1\over 8}i(2\pi\alpha')^2
\epsilon^{mnrst}C_m f_{nr}f_{st}\right)~,
\end{equation}
$x^m=\tau,x^1\dots x^4$ and ${\cal T}_4=(2\pi)^{-4}g_s^{-1}l_s^{-5}$. Upon 
compactification and expansion to second order 
the couplings to the gauge invariant operators are obtained 
\cite{Hashimoto:1998if}. We give the expression in Einstein frame:
\begin{equation}
I_{BI}={\Delta\over 4}{\cal T}_4(2\pi \alpha')^2{\rm Tr}\int {\rm d}^4x
\left({\rm e}^{-3\phi/4}\sqrt{g_{\tau\tau}}\sqrt{{\rm det}g_{\mu\nu}}f^2+
{1\over 2}i C_\tau f\tilde f\right)~.\label{bi4}
\end{equation}

{}By neglecting the metric fluctuations proportional to $k^\mu$
that give no contribution when contracted with the conserved energy momentum 
tensor of the four dimensional field theory we get the following result:
\begin{equation}
I_{BI}=I_{BI}^B+\int {\rm d}^4x \left(\psi {\cal O}_4+\chi \tilde {\cal O}_4
\right)~,
\end{equation}
where $I_{BI}^B$ accounts for the contribution from the background metric and 
dilaton fields and where we have defined the scalar and pseudoscalar glueball
operators ${\cal O}_4$ and $\tilde {\cal O}_4$ and the corresponding couplings 
$\psi$ and $\chi$ as follows:
\begin{eqnarray}
{\cal O}_4&=&{\Delta {\sqrt {f(U)}}\over 16\pi^2 g_s l_s}f_{\mu\nu}f^{\mu\nu}~,
\\
\tilde {\cal O}_4&=&{3i\Delta\over 
32\pi^2 g_s l_s}f_{\mu\nu}\tilde f^{\mu\nu}~,
\\
\psi&=&T+S~,\\
\chi&=&{3\over 2}V+{3\theta_c \over 2\pi}f \psi~.
\end{eqnarray} 
Note that the $S_4-$volume scalar fluctuation $L$ decouples.

{}The $AdS/CFT$ prescription determines the 
generating functional for the 4-dimensional field theory in terms of the low 
energy effective string theory partition function for on-shell fields
\begin{equation}
\langle{\rm e}^{\int {\rm d}^4 x \varphi_0 {\cal O}}\rangle=
e^{-I_{SG}[\varphi]}~,\label{adscft}
\end{equation}
in which the r.h.s.~containing the supergravity action $I_{SG}$ is used as an 
approximation to the string theory partition function and $\varphi_0$ are the 
boundary values of the on-shell normalizable supergravity modes $\varphi$ 
acting as sources for the field theory operator ${\cal O}$.
Therefore, to obtain the glueball spectrum we can consider the variation of 
the r.h.s.~of (\ref{adscft}) with respect to boundary sources to obtain two 
point functions. For example, for pseudoscalar glueballs we would consider 
the following variation
\begin{equation}
{\delta\over\delta V_0(x^\mu)}{\delta\over\delta V_0(y^\mu)}{\rm e}^{-I_{SG}[V,
\psi]}|_{V_0=0,\psi_0=0}=\langle\tilde{\cal O}_4(x)\tilde{\cal O}_4(y)
\rangle~.
\end{equation}
On the other hand, for scalar glueballs we find a mixing. Not only both scalars
$T$ and $S$ in the combination $\psi$ act as a source for ${\cal O}_4$ but 
also, when $\theta_c$ is nonvanishing, $\psi$ sources $\tilde{\cal O}_4$.
The corresponding two point function is obtained from the following variation:
\begin{equation}
\left({\delta\over\delta \psi_0(x)}-{\theta_c\over \pi}
{\delta\over\delta V_0(x)}\right)
\left({\delta\over\delta \psi_0(y)}-{\theta_c\over \pi}
{\delta\over\delta V_0(y)}\right)
{\rm e}^{-I_{SG}[V,
\psi]}|=\langle{\cal O}_4(x){\cal O}_4(y)\rangle~,
\label{t21}
\end{equation}
since $f(U)\rightarrow 1$ at the boundary\footnote{Eq. (\ref {t21})
makes the values $\theta=\pm \pi$ special. This indeed should be so
since one expects spontaneous $CP$ violation to take place 
for $\theta=\pm \pi$ \cite {Witten1,Witten:1998uk}. In our approach  
the form of Eq. (\ref {t21}) is an artifact 
of the normalizations that we choose to use for the bulk fields.}.

{}The above results can be summarized as follows. The field 
$\psi$ sources a linear combination of the operators 
${\cal O}_4$ and  $\tilde{\cal O}_4$, while the field 
$V$ sources only $\tilde{\cal O}_4$.  Hence, the 
diagonalization of the bulk gravity equations 
in terms of $\psi$  and $V$ suggests the 
following mixing pattern for spin-zero glueballs: 
The lowest state is a linear superposition of 
a former scalar and pseudoscalar states while the 
heavier state is intact (it is just a pseudoscalar state). 
The mass of the pseudoscalar state, determined by 
Eq. (\ref {V}), is the same as in the $\theta=0$ theory  
(determined in Refs. \cite{Hashimoto:1998if,Brower:2000rp}).
So is the mass of the mixed lightest state. Does this pattern have any  
special meaning? It seems it does. To see this consider the following 
two two-point correlation functions 
$$ C(x)\equiv \langle{\cal O}_4(x){\cal O}_4(0)\rangle -
\langle{\cal O}_4(0)\rangle ^2, $$ and 
$${\tilde C}(x) \equiv \langle\tilde{\cal O}_4(x)\tilde{\cal O}_4(0)
\rangle    - \langle\tilde{\cal O}_4(0)
\rangle^2. $$
These correlators  can be saturated by the corresponding 
orthonormal set of physical intermediate states. Then,  
the leading behavior of $C(x)$ and ${\tilde C}(x)$ at large 
euclidean $x^2$ is determined by the  corresponding lightest states. 
In a theory without the theta term those states are scalar and 
pseudoscalar glueballs 
respectively. However, once the theta term is switched on the 
scalars and pseudoscalars can mix. In general this mixing could be arbitrary.
If so, then  the former scalar state would also contribute to the expression 
for ${\tilde C}(x)$. However, because the scalar is lighter than the 
pseudoscalar (and after mixing it can only become even lighter), this 
would mean that ${\tilde C}(x)$ at large $x^2$ is completely dominated 
by the residue of the former scalar state. If this were true,  
we would not be able to determine the properties of the pseudoscalar state
by calculating ${\tilde C}(x)$. Fortunately, this does not happen.
The  contribution of the physical lightest state (which is the 
former scalar glueball) is exactly canceled in ${\tilde C}(x)$, 
as suggested by the diagonalization of the gravity equations.

%

\section*{Acknowledgments}

{}The work of AI is supported  by funds provided by New York University.

\end{document}